\documentclass[letterpaper]{article}

\usepackage{amssymb}
\usepackage{amsmath,bm}
\usepackage{float}
\usepackage{algorithm}
\usepackage{algpseudocode}
\usepackage{graphicx}
\usepackage{multirow}
\usepackage{xurl}
\usepackage[hidelinks]{hyperref}
\usepackage{xcolor}
\usepackage{bm} 
\usepackage{colortbl}
\usepackage{changes}
\usepackage{comment}
\usepackage{soul}
\usepackage{setspace}
\usepackage{arydshln}

\usepackage{graphicx}%
\usepackage{multirow}%
\usepackage{amssymb,amsfonts}%
\usepackage{amsthm}%
\usepackage{mathrsfs}%
\usepackage[title]{appendix}%
\usepackage{textcomp}%
\usepackage{manyfoot}%
\usepackage{booktabs}%
\usepackage{listings}%
\usepackage{colortbl}
\usepackage[right=1in, left=1in, top=1in, bottom=1in]{geometry} 
\usepackage[font=normalsize,labelfont=bf]{caption}
\usepackage{subcaption}
\usepackage{float} 
\usepackage{placeins} 

\usepackage{pgfplots} 

\usepackage{pdfpages}

\usepackage{afterpage} 
\usepackage[backend=biber, style=chem-acs, articletitle=true, maxnames=15]{biblatex}
\makeatletter
\def\ps@pprintTitle{%
 \let\@oddhead\@empty
 \let\@evenhead\@empty
 \def\@oddfoot{\reset@font\hfil\thepage\hfil}
 \let\@evenfoot\@oddfoot
}
\makeatother

\graphicspath{{./figures/}}   
\pgfplotsset{compat=1.18}

\usepackage{authblk}
\author[1]{Ziye Deng}
\author[1]{Reza Namakian}
\author[1,2]{Wei Gao*}
\affil[1]{J. Mike Walker $'$66 Department of Mechanical Engineering, Texas A\&M University, College Station, Texas 77843, United States}
\affil[2]{Department of Materials Science \& Engineering, Texas A\&M University, College Station, Texas 77843, United States}

\title{Amorphization-Mediated Si-I to Si-V Phase Transition and Reversible Amorphous--Si-V Phase Memory in Silicon Nanoparticles}
\date{*Email: wei.gao@tamu.edu}

\begin{document}
\maketitle

\begin{center}
\small
Published in \textit{Nano Letters}, 2026, \textbf{26}(38), 12702.\\
\href{https://pubs.acs.org/nalefd/article/26/38/12702/5429768/Amorphization-Mediated-Si-I-to-Si-V-Phase}
{Publisher version}
\end{center}

\begin{abstract}
An experiment has shown that $\sim$10-nm Si nanoparticles undergo a Si-I (diamond cubic) to Si-V (simple hexagonal) phase transition under compression, in contrast to the Si-I to Si-II (tetragonal) transition observed in bulk silicon. However, the atomistic mechanism underlying this size-dependent transition pathway remains unclear.  Here, we employ molecular dynamics simulations with a machine learning interatomic potential to reveal a stress-triaxiality-driven, two-step Si-I to Si-V transition pathway in a spherical Si nanoparticle subjected to an idealized triaxial contact loading model. An intermediate amorphous phase nucleates at the nanoparticle surface and propagates inward around the Si-I core in regions of low stress triaxiality dominated by shear. Within this amorphous shell, Si-V recrystallizes at locations with elevated stress triaxiality and hydrostatic pressure. Upon unloading, the Si-V structure transforms into an amorphous state. A subsequent loading–unloading cycle applied to this amorphous nanoparticle reveals a reversible amorphous-to-Si-V transformation, demonstrating a nanoscale phase-memory effect.

\end{abstract}





\newpage
Silicon is not only a cornerstone of modern electronics but also exhibits a rich landscape of crystalline phases, making it an ideal model system for studying stress-induced phase transitions \cite{2015_Photovoltalic_intro,MEMES_intro_1}. It is well established that the phase transition pathways in bulk silicon depend on the applied stress state and loading conditions, with the transition energy barriers modulated by the stress \cite{ghasemi2019nudged,ghasemi2020method}.
The ground state of silicon is the Si-I (diamond-cubic) structure. 
Under hydrostatic or quasi-hydrostatic pressures, such as those in diamond anvil cell (DAC) experiments, bulk silicon undergoes a well-documented sequence of phase transitions, including the Si-I$\rightarrow$Si-II (tetragonal, $\beta$-tin)$\rightarrow$Si-XI (orthorhombic, \textit{Imma})$\rightarrow$Si-V (simple hexagonal) transitions \cite{2003_DAC_review,2019_DAC_1megbar,1986_DAC_intro}. However, non-hydrostatic loading methods such as nanoindentation, which introduce localized and more pronounced deviatoric stresses, reveal different phase transition pressures and pathways~\cite{1997_expe_indent_intro,2005_Expe_Indent_intro,Non-Hydro_expe_1999}. Notably, a recent study has shown that the application of shear stress can reduce the Si-I$\rightarrow$Si-II transition pressure to as low as 0.3 GPa, far below the 16.2 GPa required under hydrostatic conditions~\cite{Levitas_2024}. In addition, bulk amorphous Si exhibits complex polyamorphic transitions and recrystallization under pressure. For example, the low-density amorphous (LDA) phase first transforms to the high-density amorphous (HDA) phase and subsequently crystallizes into Si-V~\cite{DAC_unload_amor}. Upon decompression, this Si-V phase transforms back into the LDA state. Furthermore, the LDA state converted from Si-V can reversibly transform back to Si-V, demonstrating a ``phase memory" effect in bulk amorphous silicon~\cite{porous_memory_effect}, a phenomenon also observed in materials such as AlPO4~\cite{1990_phase_memory_optic,1993_phase_memory_mechanical} and GeTe~\cite{2012_phase_memory_GeTe}.

Nanostructured Si, such as nanoparticles, nanopillars, nanocubes, and nanowires, has found widespread application in recent years owing to its unique mechanical, optical, and electrical properties. Numerous studies have shown that nanostructured silicon exhibits phase transition behaviors distinct from those observed in the bulk. For instance, an in-situ TEM compression experiment revealed a Si-I $\rightarrow$ Si-IV (hexagonal diamond) transition in 86-nm-diameter Si nanopillars when compressed along the ⟨111⟩ orientation \cite{He2016}. Conversely, another experiment on 100-nm-diameter nanopillars compressed along the [110] direction reported no phase transition even under high stress \cite{Merabet2018}. In a separate in-situ TEM study, Si nanocubes (20–65 nm) exhibited the nucleation of the Si-II phase when compressed along the [00$\overline{1}$] direction \cite{Wagner2015}.

Our study is motivated by a recent \textit{in-situ} X-ray high-pressure experiment on Si nanoparticles using DACs~\cite{PRL_Si_V}, in which no pressure-transmitting medium was employed, thereby deliberately introducing deviatoric stress into the sample. The experiment showed that $\sim$10-nm-diameter Si nanoparticles exhibited a Si-I $\rightarrow$ Si-V transition under triaxial compression, whereas larger $\sim$100-nm particles under similar loading conditions displayed a Si-I $\rightarrow$ Si-II transition, resembling bulk Si behavior. The significantly higher surface-to-volume ratio of the 10-nm particles renders surface effects much more pronounced, which can strongly dictate the phase transition pathways. 
Furthermore, the stress distribution under triaxial compression in DAC experiments is size-dependent, further influencing the transition pathways. However, owing to the limited spatial and temporal resolutions of current experimental techniques, it is challenging to isolate and understand the contributions of these factors through experiments alone.  
To address this, we employ atomistic simulations of a single 10 nm silicon nanoparticle subjected to an idealized triaxial contact loading model to investigate possible mechanisms governing the Si-I$\rightarrow$Si-V transition observed experimentally.

Density functional theory (DFT) is computationally prohibitive for studying systems at the 10-nm length scale evaluated in the previous experiment~\cite{PRL_Si_V}. As an alternative, classical molecular dynamics (MD) simulations using empirical interatomic potentials have been employed to investigate nanostructured silicon; however, these empirical potentials have yielded inconsistent results. For example, MD simulations using the Tersoff potential \cite{Tersoff_1989} suggest that Si nanoparticles (5-40 nm in diameter) subjected to uniaxial compression undergo a Si-I to Si-II phase transition \cite{Zhang2011,2007_Tersoff_beta_tin}.
By contrast, simulations using the Stillinger-Weber (SW) potential \cite{SW_1985} report only dislocation-mediated plasticity in the Si-I phase for particles of similar size (5-10 nm), providing no evidence of a phase transition \cite{Deconfinement_Plasticity}. To overcome these limitations, we employed a Gaussian Approximation Potential (GAP) developed for silicon \cite{2018_Si_GAP}. 
Machine learning interatomic potentials (MLIPs) such as GAP have reshaped computational materials science by bridging the accuracy of quantum mechanical methods with the scalability of classical MD simulations \cite{friederich2021machine,ceriotti2022beyond,shuang2025universal}. Specifically, the GAP model for silicon has demonstrated superior accuracy in capturing a wide range of structural and thermodynamic properties across diverse silicon phases \cite{2018_Si_GAP}, and it has been widely adopted to investigate both crystalline and amorphous phase transitions in bulk silicon \cite{Suggested_3_SSNEB,Suggested_2_DPMD_BulkSilicon,2024_Ge_GAP_indentation,LDA_HDA_VHDA_MD,VHDA_nature}. For comparison, we also repeated the same simulations using the classical Tersoff and SW potentials.

\begin{figure}[!t]
    \centering
    \includegraphics[width=\textwidth]
    {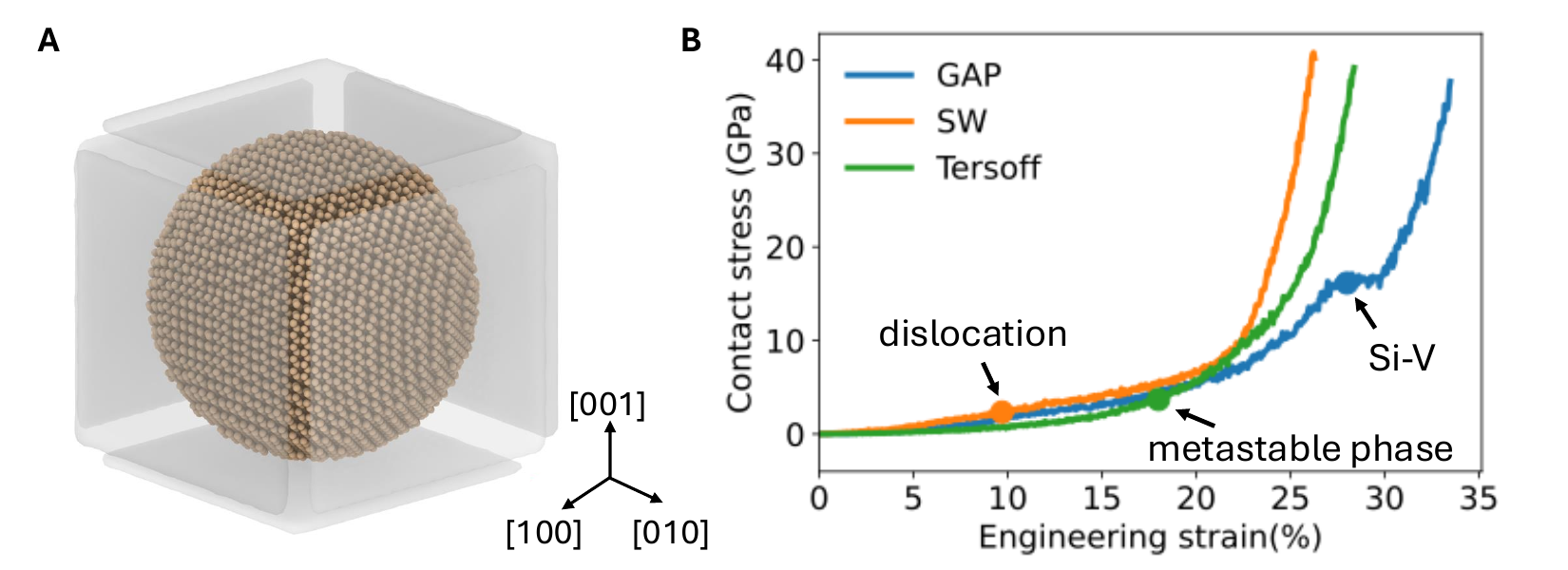}
    \caption{ (\textbf{A}) Atomistic model of a 10-nm Si nanoparticle under triaxial compression. (\textbf{B}) Stress–strain curves obtained via MD simulations using GAP, Tersoff, and SW interatomic potentials. 
    Key transition events are indicated: Si-V nucleation predicted by GAP at 28\% strain, dislocation formation predicted by SW at 9.7\% strain, and the nucleation of a metastable tetragonal phase predicted by Tersoff at 18\% strain.}
    \label{fig:compression_curves}
\end{figure}

\begin{figure}[b!]
    \centering
  \includegraphics[width=\textwidth]
  {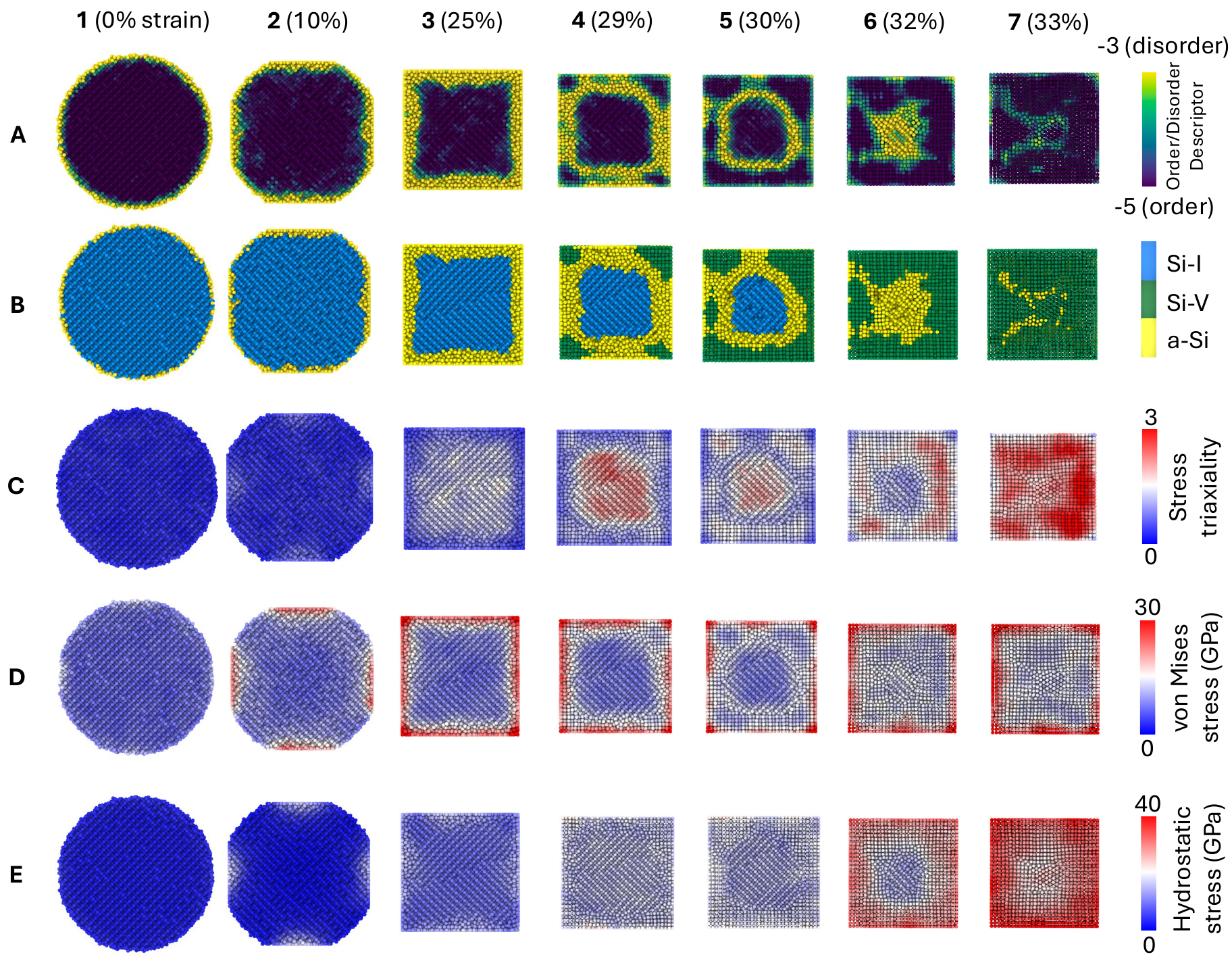}
\caption{Cross-sectional snapshots of a 10-nm Si nanoparticle under triaxial compression, displayed as a function of engineering strain (columns). The rows correspond to: (\textbf{A}) order/disorder descriptor maps; (\textbf{B}) phase classifications; (\textbf{C}) stress triaxiality fields; (\textbf{D}) von Mises (deviatoric) stress fields; and (\textbf{E}) hydrostatic stress fields.} 
\label{fig:phase_evolution}
\end{figure}

Our simulation model, illustrated in Fig.~\ref{fig:compression_curves}A, consists of a 10-nm-diameter silicon nanoparticle (26,167 atoms) in the diamond-cubic (Si-I) phase, generated using Atomsk \cite{atomsk}. 
To ensure a stable, reconstructed surface, we employed simulated annealing: the nanoparticle was first equilibrated at 800 K for 100 ps under a canonical ($NVT$) ensemble, subsequently cooled to 300 K via a 10-ps linear ramp, and finally relaxed at 300 K for 100 ps. In the recent DAC experiment~\cite{PRL_Si_V}, ensembles of Si nanoparticles were compressed in an X-ray–transparent gasket opposing diamond anvils without a pressure-transmitting medium. To approximate the highly confined non-hydrostatic conditions relevant to the DAC experiment, the nanoparticle in our MD simulations was compressed by six rigid planar indenters advancing symmetrically toward its center.

All MD simulations were performed using the LAMMPS package~\cite{LAMMPS} with a 1-fs timestep. The system temperature was maintained at 300 K via a canonical stochastic velocity rescaling (CSVR) thermostat. Notably, the CSVR thermostat yielded a $\sim$10\% increase in computational efficiency compared to the Nosé–Hoover thermostat. The indenters were modeled as repulsive surfaces having a stiffness of k=1000 eV/Å³ and a displacement rate of 0.1 Å/ps. 
To compute the per-atom virial stress tensor within the nanoparticle, atomic volumes were simultaneously evaluated using Voronoi tessellation as implemented in LAMMPS. Scalar stress metrics, including hydrostatic and von Mises stresses, were derived from the resulting stress tensor. These values were then spatially averaged over neighboring atoms within a 6~\AA~cutoff to obtain a smoothed local stress field.
Crystalline defects in the ordered regions were identified with the Dislocation Extraction Algorithm (DXA)~\cite{Ovito_DXA}. All analyses were performed using OVITO~\cite{Ovito}.

Fig.~\ref{fig:compression_curves}B compares the stress--strain responses obtained using the GAP, Tersoff, and SW potentials. The reported contact stress was calculated using the instantaneous cross-sectional area of the simulation box rather than the true atomistic contact area, which is difficult to determine accurately during the early stages of spherical nanoparticle compression. Once the nanoparticle evolves toward a cuboidal geometry, this cross-sectional area becomes a more accurate approximation of the true contact area. All three simulations exhibit a similar elastic behavior up to approximately 10\% strain, beyond which their responses diverge due to distinct underlying deformation mechanisms. In the GAP simulation, a stress plateau initiates at approximately 28\% strain and 16.1~GPa, corresponding to the nucleation and growth of the high-pressure Si-V phase. This critical stress closely matches the 14.7~GPa value reported in the recent DAC experiment~\cite{PRL_Si_V}. In contrast, the Tersoff potential predicts the nucleation of a metastable tetragonal phase, structurally related to Si-II and reversible to Si-I upon unloading, at approximately 18\% strain, followed by continued strain hardening. The SW potential predicts perfect-dislocation nucleation at approximately 10\% strain, followed by the formation of a five-coordinated body-centered tetragonal phase (BCT5) near the indenters and an abrupt slope change near 22\% strain associated with a sphere-to-cube morphological transition. Detailed SW and Tersoff results are provided in Supplementary Note~1. Notably, neither the Tersoff-type tetragonal transformation nor the SW-type dislocation-mediated plasticity was observed in the DAC experiment~\cite{PRL_Si_V}; among the three potentials, GAP best reproduces the experimentally observed Si-I$\rightarrow$Si-V transformation.

Fig.~\ref{fig:phase_evolution} presents the phase evolution predicted by GAP and elucidates the role of stress triaxiality, defined as the ratio of the hydrostatic stress to the von Mises stress. We first distinguish between ordered and disordered atoms using an entropy-based local structural fingerprint derived from the approximate two-body excess entropy expression~\cite{Ovito_Entropy}, hereafter referred to as the local order/disorder descriptor (row~A). The phase maps in row~B are then assigned based on radial distribution function (RDF) analysis, along with the angular distribution functions (ADFs) provided in Supplementary Note~2. Rows~C--E show the corresponding stress triaxiality, von Mises, and hydrostatic stress fields, respectively. During the early stages of loading, an amorphous phase nucleates at the surface and propagates inward around the Si-I core (Fig.~\ref{fig:phase_evolution}, B2--B3). 
This Si-I$\rightarrow$a-Si transition is promoted by two major factors: (i) heterogeneous nucleation at the free surface, where under-coordinated or reconstructed Si atoms facilitate bond reformation and lower the energy barrier to disordering; and (ii) a shear-dominated, relatively low stress triaxiality state adjacent to the indenters (snapshots C2-C3), accompanied by peaks in the von Mises stress (snapshots D2–D3). 
This low triaxiality results from the indenter-nanoparticle contact, which imposes a highly anisotropic stress state: the normal stress acting on the contact face is large, whereas the two in-plane principal stresses are comparatively small because the contact permits local lateral relaxation. In the absence of a pressure-transmitting medium, these stress anisotropies cannot be hydrostatically equalized; consequently, the triaxiality remains low near the contact faces even as the hydrostatic pressure increases.

With further compression to $\sim$29\% strain, coordination analysis (Fig.~S6 in the Supplementary Information) reveals that the initially formed high-density amorphous (HDA) shell (coordination number $\sim$6) densifies locally into very-high-density amorphous (VHDA) regions (coordination number $\sim$8). Subsequently, the Si-V phase nucleates from the VHDA at the four corners of the cross-section (snapshot B4 in Fig.~\ref{fig:phase_evolution}), with a corresponding critical contact stress of approximately 16 GPa. Such an a-Si$\rightarrow$Si-V transition has been observed experimentally in amorphous bulk silicon~\cite{DAC_unload_amor, porous_memory_effect, haberl2013new} and analyzed in recent GAP-based simulations~\cite{VHDA_nature}, which also identify VHDA as a precursor to the Si-V phase. In the present study, we emphasize that the nucleation is critically driven by the local stress distribution: both the stress triaxiality and hydrostatic stress rise at the corners (snapshots C4–C5 and E4–E5), demonstrating that both conditions are concurrently required. This is evidenced by the outer shell regions, which, despite experiencing high hydrostatic stresses, remain amorphous due to their low stress triaxiality (snapshots B4, C4, and E4). Meanwhile, the core region, although subjected to high stress triaxiality, retains the Si-I phase owing to its low hydrostatic stress. 
Because Si-V is considerably denser, its nucleation produces a sudden reduction in nanoparticle volume. Under displacement-controlled loading, this densification can accommodate the indenter displacement, resulting in the stress plateau in Fig.\ref{fig:compression_curves}.
As the Si-V fraction increases, its higher stiffness relative to the Si-I phase results in a hardening of the stress-strain response. A more detailed structural analysis is presented in Supplementary Note 2.

\begin{figure}[b!]
    \centering
    \includegraphics[width=\textwidth]{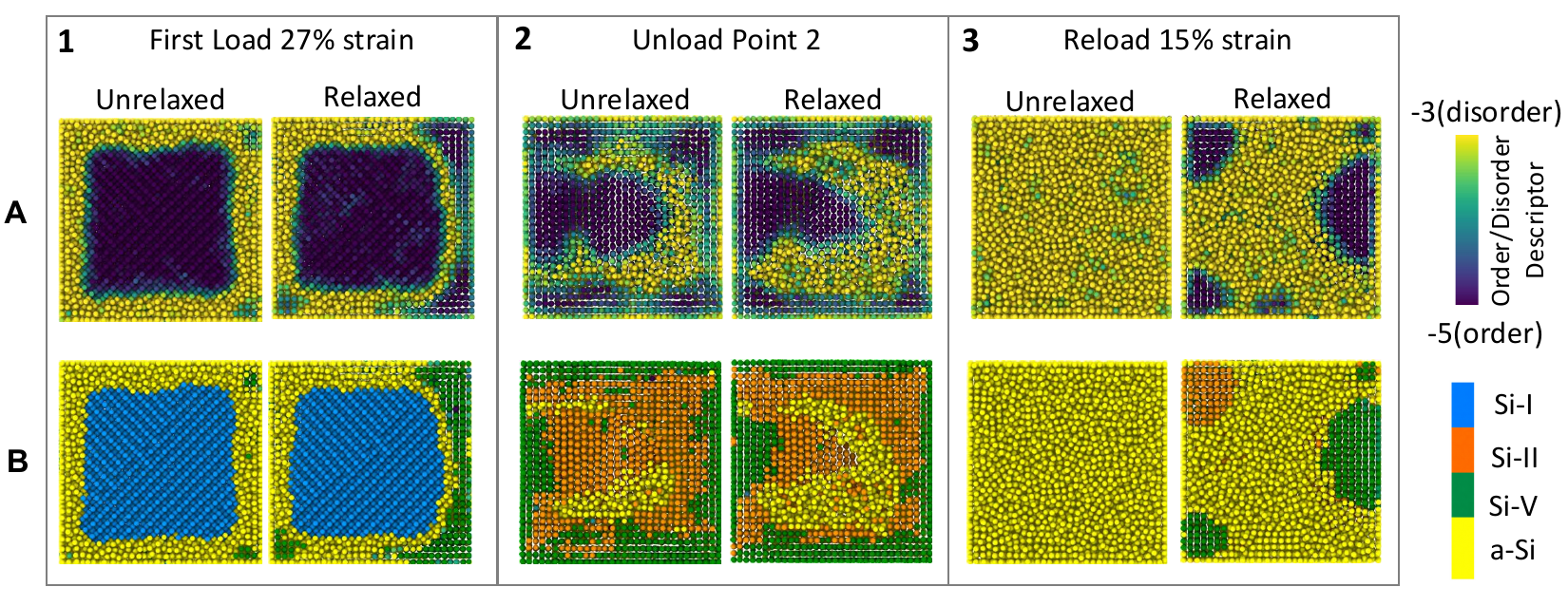}
    \caption{MD relaxation with the indenters held fixed. Configurations before and after relaxation are compared for three distinct deformation states: 27\% strain during the initial loading, critical point 2 during unloading (Fig.~\ref{fig:unload}A), and 15\% strain during reloading. (\textbf{A}) Cross-sectional snapshots colored by the order/disorder descriptor before and after relaxation. (\textbf{B}) Corresponding phase maps.}
    \label{fig:Equi}
\end{figure}
The two-step pathway described above, consisting of the Si-I$\rightarrow$a-Si transformation within the low--stress-triaxiality shell followed by the a-Si$\rightarrow$Si-V transformation in regions of simultaneously high triaxiality and high hydrostatic pressure, continues as the strain increases. At 32\% strain (Fig.~\ref{fig:phase_evolution}, B6), the Si-I core is fully consumed and replaced by the amorphous phase. With further loading, the amorphous interior recrystallizes into Si-V, whereas pockets with relatively low stress triaxiality for Si-V crystallization persist as VHDA (snapshots B7 and C7). The final transformed nanoparticle consists of multiple Si-V grains that nucleate independently within the amorphous matrix. Adjacent Si-V grains are separated either by grain boundaries or by thin amorphous layers. Although the resulting Si-I$\rightarrow$Si-V transformation agrees with the recent DAC experiment~\cite{PRL_Si_V}, that study did not report an intermediate amorphous phase. One difference between experiment and the simulation is the loading condition. Our simulation employ an idealized local contact model rather than a direct representation of the experimental powder compact. In a podwer compact, the measured XRD signal represents an ensemble average over many randomly oriented particles. To evaluate the effect of crystallographic orientation, we performed additional simulations on 10 nm silicon nanoparticles with different crystal orientations relative to the loading axis using the same approach. The results (Supplementary Note 3) show the same Si-I$\rightarrow$a-Si$\rightarrow$Si-V transformation pathway in all cases. In addition, the observed phase transition mechanism is robust with respect to variations in the indenter–nanoparticle interaction stiffness, K (Supplementary Note 4).

Another important difference between the experiment and the simulation is the loading rate. The loading rates accessible via MD are several orders of magnitude higher than those in DAC experiments, which typically operate under near-quasi-static conditions. This raises the question of whether the intermediate amorphous phase observed in MD arises from the limited timescale available for atomic rearrangement during rapid loading. Because further reduction of the loading rate in MD is computationally prohibitive, we extracted a configuration at approximately 27\% strain (just prior to the rapid growth of the high-pressure phase and near the onset of the stress plateau) and performed an additional 0.4-ns MD relaxation simulation with the indenter positions held fixed.
As shown in Fig.~\ref{fig:Equi} (panel 1), a portion of the amorphous region recrystallizes into Si-V during relaxation, whereas the majority remains amorphous (see Supplementary Note~2 and Fig.~S7 for details). The partial recrystallization of the amorphous regions into Si-V indicates that rapid MD loading can kinetically delay the a-Si$\rightarrow$Si-V transformation, allowing the amorphous phase to accumulate during compression. Concurrently, the persistence of substantial amorphous regions upon the cessation of loading suggests that the amorphous phase behaves as a metastable intermediate state within the accessible MD timescale. In DAC experiments, the much slower loading rate likely allows amorphous formation and recrystallization to occur continuously. Consequently, the amorphous state may not accumulate to an experimentally detectable volume fraction, or it may exist only locally and transiently before being consumed by Si-V growth. Thus, the absence of an experimentally reported amorphous phase does not necessarily rule out such an amorphous-mediated transformation pathway.

A related question is whether a direct crystalline Si-I$\rightarrow$Si-V transition, bypassing any intermediate phase, is mechanistically plausible. If, along a given stress path, Si-V becomes thermodynamically more stable than Si-I before Si-II or Si-XI does, then a direct Si-I$\rightarrow$Si-V transformation, in principle, could occur without passing through Si-II or Si-XI. Both the Si-I$\rightarrow$a-Si and Si-I$\rightarrow$Si-V (if it exists) transitions are expected to proceed through nucleation and growth, rather than a concerted transformation in which all atoms move collectively. Our MD results suggest that the amorphous-mediated pathway is kinetically accessible on the simulation timescale, but they do not exclude the possibility of a direct crystalline transformation under different conditions. Quantitatively comparing the barriers for the amorphous and Si-V nucleation within the Si-I matrix would require large-scale saddle point searches and is a valuable direction for future work.

In addition to the factors discussed above, the limitations of the interatomic potential itself should also be considered.  Although the GAP training database includes most relevant crystalline phases, amorphous structures, and various surface reconstructions, this does not guarantee that all predicted phenomena are fully realistic. Like all machine-learning interatomic potentials, the predictive capability of GAP depends on the coverage of the training database, and configurations that differ substantially from those represented in the training set may be subject to extrapolation errors. For example, a recent study on low-pressure phase transformation in amorphous silicon showed that the GAP potential is not necessarily superior for describing all metastable silicon phases \cite{Haberl_r8_2025}.

\begin{figure}[b!]
    \centering
    \includegraphics[width=\textwidth]{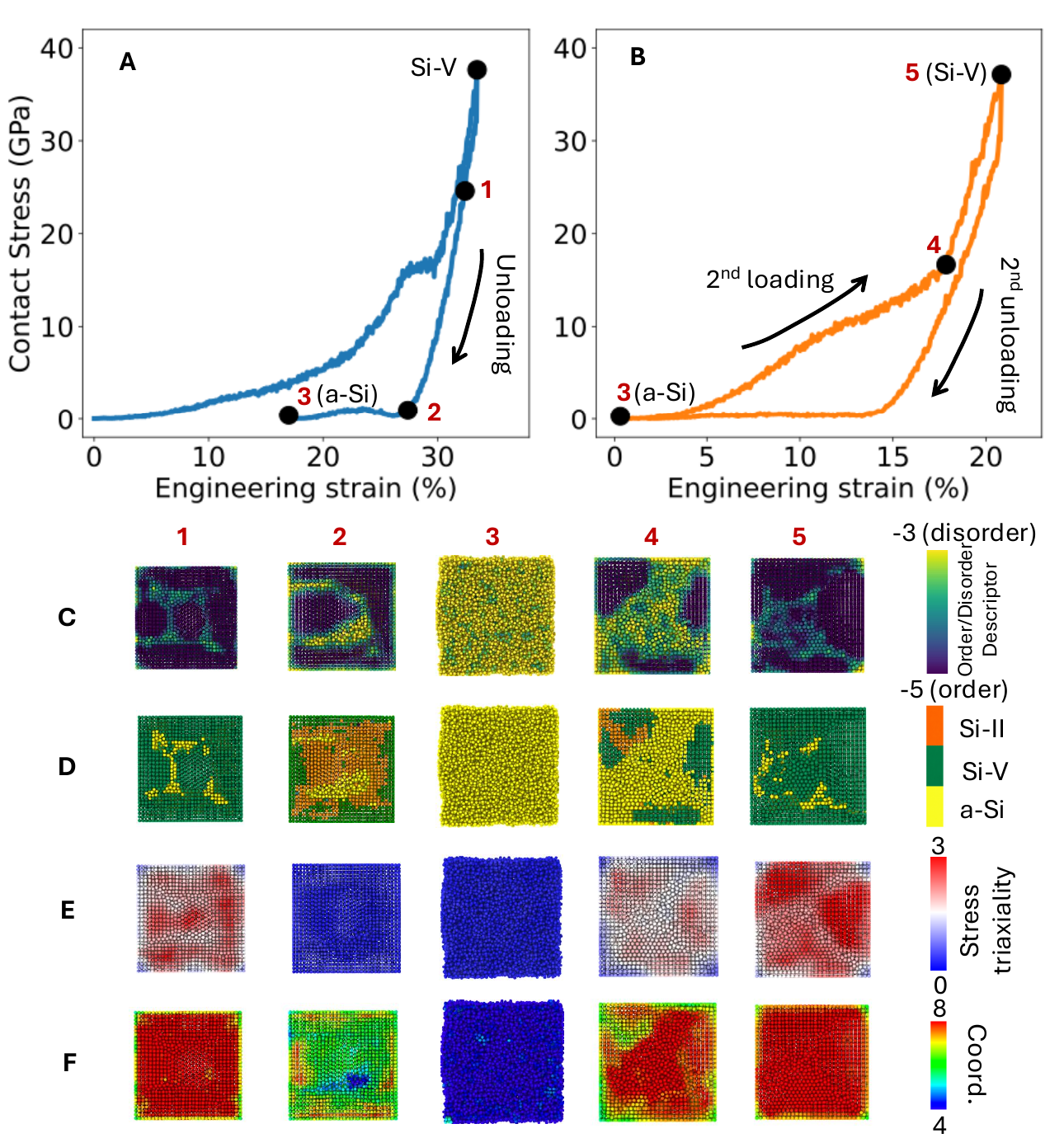}
\caption{(\textbf{A}) Stress–strain responses obtained from triaxial compression and unloading MD simulations using the GAP interatomic potential. The trajectory reveals that the a-Si intermediate phase mediates the Si-I$\rightarrow$Si-V transition, whereas unloading leaves the nanoparticle predominantly a-Si. (\textbf{B}) A subsequent loading–unloading cycle demonstrates a reversible a-Si$\leftrightarrow$Si-V transition. (\textbf{C-F}) Cross-sectional snapshots of the Si nanoparticle at representative points along the stress–strain curves.}
    \label{fig:unload}
\end{figure}

To check whether the observed transformation mechanism persists at smaller particle sizes, we additionally simulated a 5 nm diameter nanoparticle using the same simulation protocol. The results are shown in Fig.~S20 in Supplementary Note 6. The 5 nm nanoparticle follows the same two-step transformation pathway, Si-I$\rightarrow$a-Si$\rightarrow$Si-V. Compared with the 10 nm particle, reducing the particle size leads to a larger fraction of amorphous silicon, caused by a stronger influence of the free surface. It is also worth noting that previous first-principles MD simulations \cite{1.9nm_DFT} identified amorphization as the dominant deformation mechanism in 1.9 nm silicon nanoparticles under compression, which is consistent with the trend observed here. On the other hand, previous work showed that 30 nm silicon nanoparticles exhibit different transformation pathways, namely Si-I$\rightarrow$Si-XI$\rightarrow$Si-V under hydrostatic loading and Si-I$\rightarrow$Si-II$\rightarrow$Si-XI$\rightarrow$Si-V under non-hydrostatic loading \cite{Levitas_2024}. This suggests that a mechanism change could occur between 10 and 30 nm. Unfortunately, extending the present GAP-based MD simulations to 30 nm is computationally prohibitive. Employing a more computationally efficient MLIP, such as the neuroevolution potential (NEP) \cite{fanNeuroevolutionMachineLearning2021} implemented in GPUMD \cite{fanGPUMDPackageConstructing2022}, may provide a practical way for studies in this size range.

We next examine the unloading behavior and the subsequent loading–unloading cycle. The stress–strain curve during the initial unloading path (Fig.~\ref{fig:unload}A) exhibits a steep initial drop, reflecting the high stiffness of the Si-V-rich state. In regions of low stress triaxiality (corresponding to high deviatoric stress), an intermediate Si-II phase nucleates within the Si-V matrix (Fig.~\ref{fig:unload}, snapshot D2), as confirmed by RDF and ADF analyses detailed in Supplementary Note 5. Although the Si-II phase is typically regarded as a high-pressure phase in bulk silicon, recent work has shown that it can also be stabilized at low pressures in the presence of high deviatoric stresses~\cite{Levitas_2024}. In our case, however, the Si-II phase is transient: it rapidly transforms into the amorphous phase even during relaxation with the indenters held fixed (Fig.~\ref{fig:Equi}, panel 2). Beyond critical point 2 in Fig.~\ref{fig:unload}A, continued amorphization is accompanied by volumetric expansion, which generates a transformation backstress,  leading to a slight increase in the reaction stress. Upon complete unloading, the nanoparticle relaxes into a low-density amorphous (LDA) state characterized by a predominantly fourfold-coordinated network (Fig.~\ref{fig:unload}, D3 and F3).

The subsequent cycle, initiated from the recovered amorphous state, demonstrates a reversible a-Si$\leftrightarrow$Si-V transformation (Fig.~\ref{fig:unload}B). During the second loading path, the tangent stiffness increases sharply near critical point~4, coincident with the densification of LDA/HDA into VHDA within low-stress-triaxiality regions and the nucleation of ordered domains in regions of high stress triaxiality (Fig.~\ref{fig:unload}, snapshots D4--E4). 
Most of these ordered domains are identified as the Si-V phase, although small accompanying Si-II regions are also observed, producing mixed Si-II/Si-V ordered domains. This mixed-phase character is also observed during the fixed-indenter relaxation of the reloading configuration at 15\% strain (Fig.~\ref{fig:Equi}, panel 3). Following relaxation at fixed indenter positions, ordered domains emerge from the amorphous matrix, and the relaxed phase map exhibits localized coexistence of the Si-II and Si-V order. Similar behavior has been reported in bulk amorphous Si, in which HDA relaxation near 14~GPa crystallized toward Si-V via a Si-II-mediated pathway~\cite{LDA_HDA_VHDA_MD}.

By the end of the second loading, the nanoparticle is predominantly Si-V, with only small VHDA pockets persisting in low-triaxiality regions, similar to the first loading path. The subsequent unloading returns the structure to a predominantly amorphous state and, unlike the first cycle, does not introduce additional residual strain. This reversible amorphous--crystalline transformation ``memory'' has been reported in bulk porous Si under compression--decompression cycles~\cite{porous_memory_effect}. Here, we demonstrate an analogous nanoparticle-scale memory effect: the first loading cycle prepares an amorphous precursor from the initial Si-I nanoparticle, enabling repeatable a-Si$\leftrightarrow$Si-V transformations during subsequent cycles. It is noted that the transformation pathway of amorphous silicon is likely sensitive to initial amorphous structure, sample geometry and loading conditions. Therefore, different pathways may emerge under different conditions. For example, a recent study reported direct nucleation of the metastable r8 phase from bulk amorphous Si and Ge under compression below 10~GPa \cite{Haberl_r8_2025}.

In conclusion, MD simulations using the GAP interatomic potential uncover a previously unrecognized two-step phase-transition pathway in 10-nm silicon nanoparticles under triaxial compression. Specifically, a metastable amorphous shell initially forms around the Si-I core in regions of low stress triaxiality. This amorphous phase subsequently densifies and recrystallizes into the Si-V phase in regions where elevated hydrostatic stress and elevated stress triaxiality coexist.  The observed Si-I$\rightarrow$Si-V transformation is consistent with recent DAC experimental observations on nanoparticles of a similar size.
Upon unloading, the Si-V–dominated nanoparticle transforms into a fully amorphous state. Furthermore, a subsequent loading cycle applied to this a-Si nanoparticle reveals a reversible a-Si$\rightarrow$Si-V transition, whereas a final unloading path returns the nanoparticle to its amorphous state. Finally, our study illustrates the combined roles of hydrostatic stress and stress triaxiality in the Si-I$\rightarrow$a-Si$\rightarrow$Si-V transformation. An important extension of the present work is to establish a quantitative transformation criterion under systematically controlled stress states, analogous to the method proposed by Levitas et al.~\cite{PRB_Levitas_2017}. This may provide an analytical framework for understanding surface-assisted amorphization and the subsequent Si-V nucleation in silicon nanoparticles.

\section*{Supporting Information}
The Supporting Information is available free of charge.
\begin{itemize}
    \item Supplementary Note 1: Phase Transition of 10 nm Si Particle under Triaxial
    Compression Predicted by SW and Tersoff Potentials
    \item Supplementary Note 2: Structural Analysis of Si-I
    $\rightarrow$ HDA $\rightarrow$ VHDA $\rightarrow$ Si-V during Initial Loading
    \item Supplementary Note 3: Simulations for Nanoparticles with Different Crystallographic Orientations
    \item Supplementary Note 4: Effects  of Wall--Atom Interaction Stiffness $K$
    \item Supplementary Note 5: Polymorphic Transitions during Unloading
    from the Si-V-rich State
    \item Supplementary Note 6: Transformation Pathway in a 5-nm Si Nanoparticle
    \item Supplementary Note 7: Initial Surface Structure Analysis
\end{itemize}

\section*{Data availability statement}
The MD simulation data files, input scripts, and simulation movies are available on GitHub at: \href{https://github.com/Gao-Group/Silicon_Triaxial_MD}{Gao-Group/Silicon\_Triaxial\_MD}.

\section*{CRediT authorship contribution statement}
\textbf{Ziye Deng}: Methodology, Software, Investigation, Writing - original draft, Writing - review \& editing. \textbf{Reza Namakian}: Methodology, Software, Investigation, Writing - review \& editing. \textbf{Wei Gao}: Conceptualization, Methodology, Software, Investigation, Writing - original draft, Writing - review \& editing, Supervision, Funding acquisition.

\section*{Notes}
The authors declare no competing financial interest.

\section*{Acknowledgments}
W.G. thanks Prof. Albert Bart\'{o}k from the University of Warwick, UK for helpful discussions regarding the GAP potential. W.G. gratefully acknowledges financial support of this work from the National Science Foundation under Grant no.
CMMI-2305529. The authors acknowledge the Texas Advanced Computing Center (TACC) at The University of Texas at Austin and Texas A\&M High Performance Research Computing for providing the computing resources that contributed to the research results reported within this paper.

\printbibliography

\end{document}